\documentclass[aps,prl,reprint,superscriptaddress]{revtex4-1}%preprint
\usepackage{graphicx}% Include figure files
\usepackage[usenames,dvipsnames]{xcolor}
\usepackage{amssymb,amsmath}
\usepackage[breaklinks]{hyperref}

\graphicspath{{images/}}
\begin{document}

\newcommand{\CeCoIn}{CeCoIn$_5$}
\newcommand{\muSR}{$\mu$SR}
\newcommand{\Tc}{$T_{c}$}
\newcommand{\modif}[1]{\textcolor{red}{#1}}

\title{Strong pressure dependence of the magnetic penetration depth in single crystals of the heavy fermion superconductor CeCoIn$_5$ studied by muon spin rotation} 
\author{L.~Howald}
\email[]{ludovic.howald@physik.uzh.ch}
%\thanks{}
\affiliation{Physik-Institut der Universit\"at Z\"urich, Winterthurerstrasse 190, CH-8057 Z\"urich, Switzerland}
\author{A.~Maisuradze}
\affiliation{Physik-Institut der Universit\"at Z\"urich, Winterthurerstrasse 190, CH-8057 Z\"urich, Switzerland}
\affiliation{Laboratory for Muon Spin Spectroscopy, Paul Scherrer Institut, CH-5232 Villigen PSI, Switzerland}
%\author{Z.~Guguchia}
%\affiliation{Physik-Institut der Universit\"at Z\"urich, Winterthurerstrasse 190, CH-8057 Z\"urich, Switzerland}
\author{P.~Dalmas de R\'eotier}
\affiliation{Institut Nanosciences et Cryog\'enie, SPSMS, CEA and University Joseph Fourier, F-38054 Grenoble, France}
\author{A.~Yaouanc}
\affiliation{Laboratory for Muon Spin Spectroscopy, Paul Scherrer Institut, CH-5232 Villigen PSI, Switzerland}
\affiliation{Institut Nanosciences et Cryog\'enie, SPSMS, CEA and University Joseph Fourier, F-38054 Grenoble, France}
\author{C.~Baines}
\affiliation{Laboratory for Muon Spin Spectroscopy, Paul Scherrer Institut, CH-5232 Villigen PSI, Switzerland}
\author{G.~Lapertot}
\author{K.~Mony}
\author{J.-P.~Brison}
\affiliation{Institut Nanosciences et Cryog\'enie, SPSMS, CEA and University Joseph Fourier, F-38054 Grenoble, France}
\author{H.~Keller}
\affiliation{Physik-Institut der Universit\"at Z\"urich, Winterthurerstrasse 190, CH-8057 Z\"urich, Switzerland}

\date{\today}
~\\
~\\
\begin{abstract}
In the tetragonal heavy fermion system CeCoIn$_5$ the unconventional superconducting state is probed by means of muon spin rotation.
The pressure dependence ($0-1$~GPa) of the basal-plane magnetic penetration depth ($\lambda_a$), the penetration depth anisotropy ($\gamma=\lambda_c/\lambda_a$) and the temperature dependence of $1/\lambda^2_i$ ($i=a,c$) were studied in single crystals.
A strong decrease of $\lambda_a$ with pressure was observed, while $\gamma$ and $\lambda_i^2(0)/\lambda_i^2(T)$ are pressure independent. A linear relationship between $1/\lambda_a^2(270\text{~mK})$ and \Tc{} was also found. 
The large decrease of $\lambda_a$ with pressure is the signature of an increase of the number of superconducting quasiparticles by a factor of about 2.
\end{abstract}

%74.70.Tx 	Heavy-fermion superconductors
%75.30.Mb 	Valence fluctuation, Kondo lattice, and heavy-fermion phenomena (see also 71.27.+a Strongly correlated electron systems, heavy fermions; for heavy-fermion superconductors, see 74.70.Tx)
%71.27.+a Strongly correlated electron systems, heavy fermions)
%76.75.+i 	Muon spin rotation and relaxation
%74.62.Fj 	Effects of pressure

\maketitle

Unconventional superconductors are characterized by their proximity to different instabilities. In heavy fermion systems superconductivity  is often found in the region of the phase diagram where a weak magnetic phase disappears 
\cite{Mathur1998}. However, in some systems \cite{Jaccard1999} superconductivity is also detected in proximity of a valence phase transition.

\CeCoIn{} is a prototypal heavy fermion superconductor \cite{petrovic2001} at the focus of numerous studies owing to the proximity of quantum criticality. This proximity is reflected by the pronounced non-Fermi liquid features \cite{Sidorov2002, *Nakajima2007} and the highest superconducting (SC) transition temperature $T_c = 2.3$~K \cite{petrovic2001} among the Ce based heavy fermions. In addition, this tetragonal system is characterized by a quasi two-dimensional Fermi surface \cite{Settai2001,Maehira2003} and a two-gap \cite{Seyfarth2008} unconventional SC state with d-wave  symmetry \cite{Izawa2001,*Aoki2004}.

In order to clarify the relation between the SC phase and quantum criticality, the evolution of basic SC parameters with respect to a tuning parameter is required. In this letter the pressure (p) and temperature (T) dependence of a fundamental SC quantity --- the magnetic penetration depth ($\lambda_i$) --- was studied. 
Here $i=a,c$ corresponds to a screening current flowing along the main crystallographic directions: perpendicular, respectively parallel to the $c$-axis.
$\lambda_i$ is obtained through the precise magnetic field distribution in the SC vortex state probed by transverse-field (TF) muon-spin rotation ($\mu$SR).

\CeCoIn{} is in the clean limit with a ratio of coherence length to mean free path $\xi/l<0.02$ ($\xi<8.2$~nm) \cite{Settai2001,Seyfarth2008}. In this limit the penetration depth can be written in the London model as:
\begin{equation}\label{London}
1/\lambda_i^2 = \mu_0 e^2 n_S/m_i^\star 
\end{equation}
Here $\mu_0$ is the vacuum permeability, $e$ the electron charge, $n_S$ the number density (number of superconducting quasiparticles), and $m_i^\star$ the effective quasiparticle mass.

The \muSR{} experiments were performed at the Swiss Muon Source (S$\mu$S), Paul Scherrer
Institute (PSI), Switzerland, using the GPD (under $p$) and LTF (ambient $p$, low $T$) spectrometers. 
In a TF-\muSR{} experiment spin polarized positive 
muons are implanted into a sample in an external magnetic field $\mu_{0}H$ (field cooled from above \Tc ) applied perpendicular to the initial muon-spin polarization. %The momentum of the muons can be tuned to adjust the depth of implantation \cite{Maisuradze2011}. 
In the presence of a magnetic field at the muon site $B_{\mu}$ %(external and/or internal magnetic field) 
the muon spin precesses at its Larmor frequency $\omega_{\mu} = \gamma_{\mu}B_{\mu}$ ($\gamma_{\mu}=8.516\cdot 10^8$~rad~s$^{-1}$T$^{-1}$ is the gyromagnetic ratio of the muon) before decaying with a life time of $\tau_\mu=2.2$~$\mu$s into a positron and two neutrinos. 
Due to parity violation the decay positron is preferentially emitted along the muon spin direction. 
Forward and backward positron detectors with respect to the initial muon polarization are used to monitor the \muSR{} asymmetry spectrum $A(t)$.

\begin{figure}
\includegraphics[width=0.48\textwidth]{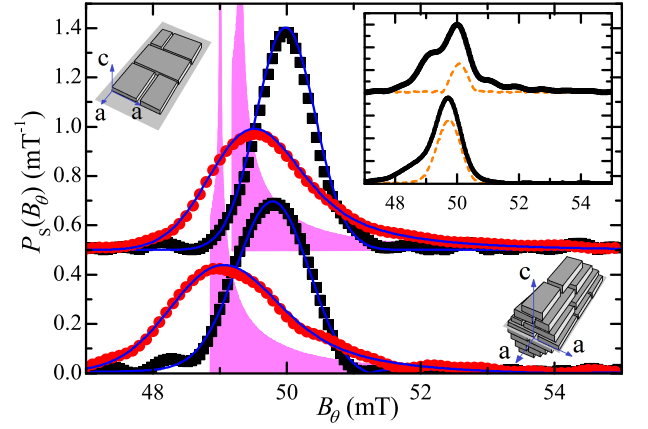}%
\caption{\label{FT}(Color online) Local magnetic field distributions $P_S(B_\theta,T)$ %(background subtraction described in text) 
of \CeCoIn{} at $\mu_0H\approx 50$~mT for $H\parallel c$ in the normal (black squares) and SC (red circles) state obtained with the GPD (bottom; $p=0.2$~GPa, $T \simeq 270$~mK) and the LTF (top; $p=0$~GPa, $T \simeq 19$~mK) spectrometers. The blue lines are FT of $R_{S}(t)$. The magenta area represent $0.5 P_{VL}(B_\theta,T)$. 
 For clarity, the LTF field distributions are shifted vertically. A Gaussian apodization ($A(t)\, \exp{[-1/2(t/\tau)^2]}$) of $\tau=$4 $\mu$s (5 $\mu$s) was used for the GPD (LTF) spectra.
 In the inset the FT of $A(t)$ (black lines) is shown. The orange dotted lines are the FT of $R_{Bg}(t)$.
 %correspond to the field distribution originating from the background signal: the FT of $R_{Bg}(t)$.
 } 
\end{figure}

Single crystals of \CeCoIn{} were grown by indium flux method \cite{Canfield1992} (rare earth from \cite{MPC}), centrifuged and etched in HCl solution to remove the indium excess. Thin plate-like single crystals were obtained with their large faces corresponding to the (001) basal plane. 
 Using this particular geometry, two samples were prepared, consisting of $\sim$10, respectively $\sim$200 crystals glued together with G.E. varnish, as sketched in Fig.~\ref{FT}.
The mosaic sample ($c$-axis normal to the plane) was studied with the LTF spectrometer. The cylindrical-like sample ($a$-axis is the main axis) was mounted in a piston cylinder pressure cell of CuBe alloy with Daphne oil as a pressure transmission medium \cite{Maisuradze2011} and measured with the GPD spectrometer. The actual pressure in the cell was determined by the \Tc{} of a small piece of indium.

For different pressures an angular scan consisting of 5-8 TF-\muSR{} spectra was taken at $T \simeq 270$~mK with an applied field $\mu_0H \simeq 50$~mT forming an angle $\theta$ with the sample's crystallographic $c$-axis. For $p =$ 0~GPa, 0.2~GPa, and 0.6~GPa a temperature scan was also recorded for $\theta=0^\circ$ ($H\parallel c$) and $\theta=90^\circ$ ($H\perp c$). 
The field was chosen to be higher than the critical field of bulk indium ($\mu_0H_{c2}(0)=23$~mT) to avoid artifacts due to possible residual flux from the growth. For comparison the values of the Pauli limited critical fields for \CeCoIn{} are: $\mu_0H_{c1}\simeq 10$~mT ($H\parallel c$ and $H\perp c$) \cite{Seyfarth2008}, $\mu_0H_{c2}=5$~T ($H\parallel c$) and 11.5~T ($H\parallel c$) \cite{Settai2001}. 
A field of $\mu_0H=50$~mT is also small enough so that the Knight shift \cite{Higemoto2010} and Zeeman current \cite{Spehling2009,*Dalmas2011} effects can be neglected in the analysis of the spectra.
In the normal state about $6\cdot 10^6$ and in the SC state $10-20\cdot 10^6$ positrons events were recorded for a \muSR{} time spectrum. 

%At several pressures zero field and longitudinal field (0-0.6 T) \muSR{} measurements were also performed. As for ambient pressure \cite{Higemoto2002}, no indication of long-range magnetic order was found.  
%Similarly, no significant temperature dependence of the signal was observed, implying that no magnetic fluctuations are present. The measured damping rate is in the range expected for a field distribution at the muon sites of nuclear origin only.

\begin{figure}
\includegraphics[width=0.48\textwidth]{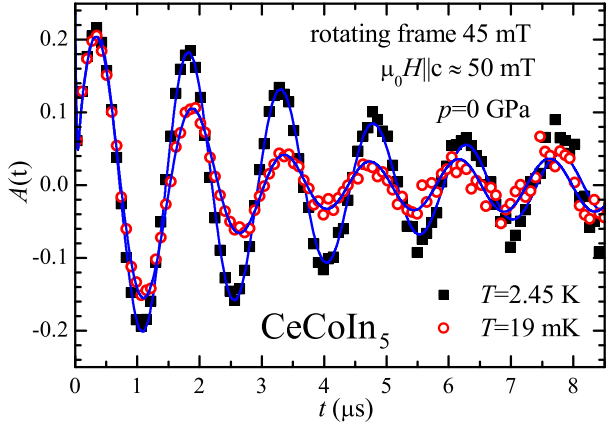}%
\caption{\label{Signal}(Color online) \muSR{} asymmetry spectrum $A(t)$ of \CeCoIn{} at ambient pressure in the normal (black squares) and SC state (red circles) taken with the LTF spectrometer at $\mu_0H=50$~mT for $H\parallel c$. %For a better visualization 
The data are shown in a frame rotating at a frequency corresponding to $45$~mT \cite{Yaouanc2011}.  
Solid lines represent fits to the data (see text).
} 
\end{figure}

Typical TF-$\mu$SR time spectra in the normal and SC state are displayed in Fig.~\ref{Signal}. The temporal oscillations and damping of the $\mu$SR asymmetry reflect directly the local magnetic field distribution at the muon stopping sites. 
The $\mu$SR time spectrum consists of two contributions: a background signal ($Bg$) arising from the muons stopping in the silver sample holder for the LTF spectrometer or the pressure cell for the GPD spectrometer and a signal arising from the muons stopping in the sample ($S$) \cite{Maisuradze2011}.
These contributions are clearly seen in Fig.~\ref{Signal}. At short times the sample contribution dominates: in the SC state the damping of the signal is enhanced due to the field broadening generated by the vortex lattice (VL), and the oscillating frequency is reduced due to diamagnetic screening. In contrary, for $t > 7\, \mu$s in the normal state and $t > 3.5\, \mu$s in the SC state, only the signal of the muons stopping in the silver background persists.
The \muSR{} time spectra are well described with the following equation:
\begin{equation}\label{At}
A(t)=A_{0}[(1-f_S(\theta))R_{Bg}(t)+f_S(\theta)R_{S}(t)]
\end{equation}
Here $f_S(\theta)$ denotes the fraction of muons stopping in the sample, $A_{0}$ the initial asymmetry of the signal and $R_{S}(t)$ [$R_{Bg}(t)$] is the sample [background] muon depolarization function. $f_S(\theta)$ was determined to be $\simeq 82\%$ for the LTF spectrometer and typically $\simeq 45\%$ for the GPD spectrometer.
Here $f_S(\theta)$ varies each time the pressure cell is manipulated (change of $p$ or $\theta$) as the sample position relative to the muon beam is modified.
In various configurations we recorded a \muSR{} spectrum after a small field increase of 4~mT at low temperatures. Due to pinning the shift of field in the sample is less, allowing to determine precisely the fraction of muons stopping in the sample.
The background depolarization function is described by a Gaussian field distribution \cite{Maisuradze2011}:
\begin{equation}\label{RBgt}
R_{Bg}(t)=\cos(\gamma_\mu \langle B_{Bg}(\theta,T) \rangle t+\phi_0)\exp({-\gamma_\mu^2\sigma_{Bg}^2(\theta,T)t^2}/2)
\end{equation}
The average background magnetic field 
$\langle B_{Bg}(\theta,T) \rangle\simeq 50$~mT and the standard deviation
of the Gaussian field distribution $\sigma_{Bg}(\theta,T)$ vary in the SC state since the diamagnetic sample induces a field inhomogeneity in its surrounding. The initial phase $\phi_0$
%, due to the precession of the muon spin prior its implantation, 
 is constant. 

The sample depolarization function may be written as:
\begin{equation}\label{RSt}
	\begin{array}{r l}
	R_{S}(t)=&\exp({-\gamma_\mu^2\sigma_{S}^2(\theta,T)t^2/2})  \\
	 &\times \int{P_{VL}(B_\theta,T)\cos(\gamma_\mu B_\theta t+\phi_0) \, dB_\theta}\\
	\end{array}
\end{equation}
The presence of a VL gives rise to a local magnetic field distribution along the direction of the applied field $P_{VL}(B_\theta,T)$, reflected by the integral in Eq.~(\ref{RSt}). For an extreme type-II superconductor in the London limit $P_{VL}(B_\theta,T)$ is uniquely determined by an effective penetration depth $\lambda_{eff}(\theta,T)$ \cite{Brandt1988b,*Maisuradze2009,Yaouanc2011}. For the two principal magnetic field orientations one has: $\lambda_{eff}(\theta=0^\circ,T)=\lambda_a(T)$ and $\lambda_{eff}(\theta=90^\circ ,T)=\sqrt{\lambda_a(T) \lambda_c(T)}$. 
The first factor in Eq.~(\ref{RSt}) describes the muon depolarization due to additional contributions ($\sigma_{S}^2(\theta,T)=\sigma_N^2(\theta)+\sigma_{dVL}^2(\theta,T)$)\footnote{
%The presence of two different muon stopping sites with different Knight shifts gives rise to an additional broadening $\sigma_K \approx 0.2$~mT for an applied field of 50~mT.  This broadening which is reduced by nearly a factor 2 \cite{Higemoto2010} in the SC state is negligible compared to the other contributions.
The presence of two different muon stopping sites with different Knight shifts produces a broadening proportional to the applied field: $\sigma_K \approx 0.2$~mT at 50~mT in the normal phase. This broadening which is reduced by a factor nearly 2 \cite{Higemoto2010} in the SC state is negligible at 50~mT compared to the other contributions, but it is not for sizably larger fields.
}: (i) the nuclear moments [$\sigma_N(\theta)\approx 0.5$~mT] and (ii) the disorder of the VL [$\sigma_{dVL}(\theta,T)=\sigma_{dVL0}(\theta)\lambda_{eff}^2(\theta,0)/\lambda_{eff}^2(\theta,T)$ with $\sigma_{dVL0}(\theta)\approx 0.5$~mT \cite{Riseman1995}]. The local magnetic field distribution in the sample $P_{S}(B_\theta,T)$ can be obtained 
from the cosine Fourier transformation (FT) of the experimentally measured $A(t)$, after subtraction of
 %the background signal 
 $R_{Bg}(t)$ (Fig.~\ref{FT}).

\begin{figure}
\includegraphics[width=0.48\textwidth]{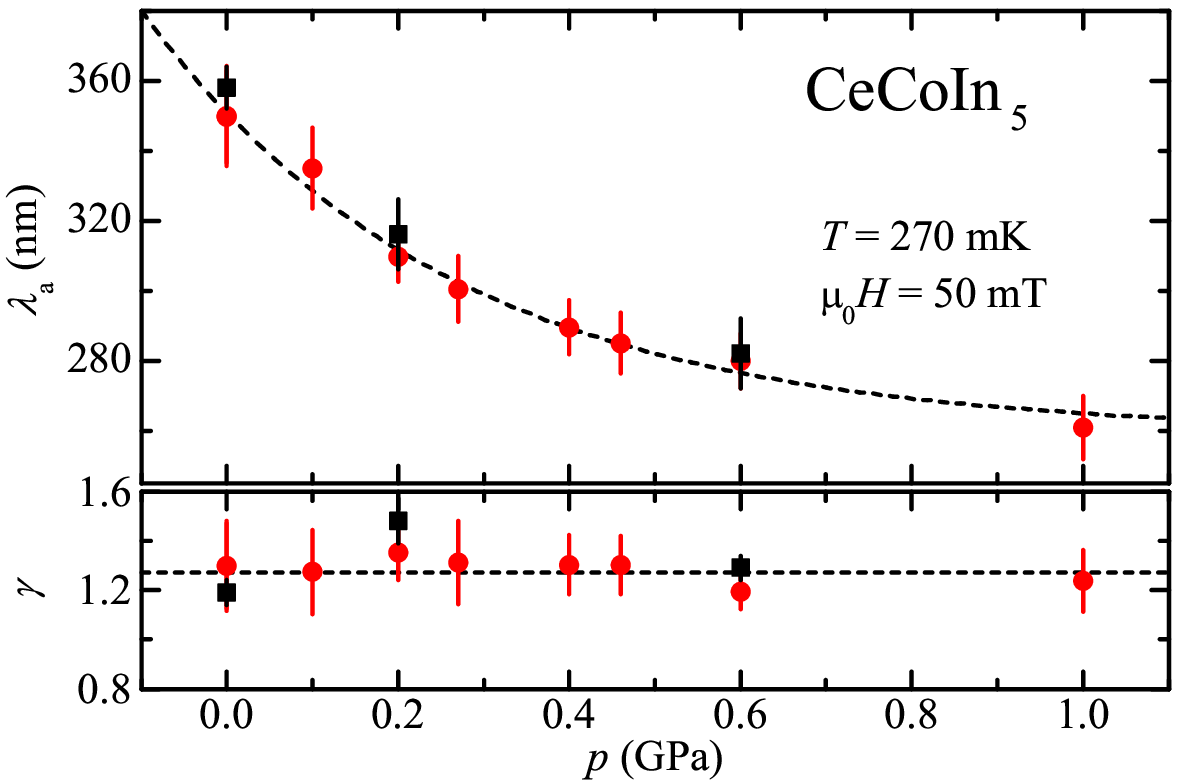}%
\caption{\label{Lambda_a}(Color online) Pressure dependence of $\lambda_a$ and $\gamma=\lambda_c/\lambda_a$ in \CeCoIn . Red circles are obtained from angular dependent spectra, black squares from the temperature dependent ones. Dashed lines are guides for the eyes.}
\end{figure}

The angular dependent spectra were analyzed globally [Eqs.~(\ref{At}) to (\ref{RSt})] \footnote{\label{globally} Explicitly, this means that all the spectra taken at a given $p$ and $T$ for different $\theta$ or at a given $p$ and $\theta$ for different $T$ are fitted together with the constraints given in the text. Each point of the asymmetry spectrum $A(t)$ is weighted statistically. } with the constraint that $\lambda_{eff} (\theta,T)=\lambda_a(T) \sqrt{\cos^2(\theta)+\gamma(T) \sin^2(\theta)}$ with $\gamma(T)=\lambda_c(T)/\lambda_a(T)$  \cite{Bulaevskii1992}. $\lambda_{eff} (\theta,T=270\text{~mK})$ was used to determine the exact orientation of the sample in the pressure cell (position of $\theta=0^\circ$).
The pressure dependence of the obtained parameters $\lambda_a$ and $\gamma$ are shown in Fig.~\ref{Lambda_a} (red circles). 
The analysis of the temperature dependent spectra treated globally \cite{Note2} is also presented (black squares).
The two data sets give similar results, although different assumptions were made ($T$ or $\theta$ dependence of some parameters fixed), demonstrating the reliability of the model.

The analysis yields at ambient pressure a value of $\lambda_a(T\rightarrow 0{\rm K},\mu_0H=50 {\rm mT})= 350(12)$~nm. Since in the London model the contribution of the vortex core is neglected, this value is overestimated \cite{Yaouanc1997}. Taking $\mu_0H_{c2\, Orb.}\simeq 7.5$~T for the orbital upper critical field 
%from the model of $H_{c2}(T,p)$ developed in 
\cite{Howald2011b}, this correction is only $\simeq 4$\% for an applied field of $\mu_0H=50$~mT and therefore was neglected.
In comparison, the first \muSR{} experiment reported $\lambda_{a}(0) \approx 550$~nm \cite{Higemoto2002}. 
This experiment was performed in a large magnetic field $\mu_0H = 0.3$~T, and in the analysis an additional field broadening was neglected \cite{Note1}. This is very likely the main reason for the larger value of $\lambda_a(0)$.
Neutron diffraction experiments reported $\lambda_{a}(0)$ between 247(10)~nm \cite{Eskildsen2003} and $\simeq 465$~nm \cite{Bianchi2008}. The first value, measured in a magnetic field of 2~T, was underestimated because Zeeman currents that produce an additional contribution to the field broadening \cite{Michal2010} were neglected. The second value was deduced from measurements at 0.5~T. Including the correction for the vortex cores \cite{Yaouanc1997}, we obtain $\lambda_a(0)\approx 360$~nm in agreement with the present value. Surface impedance techniques provided smaller  %$\lambda_a(0)= \sim 190$ nm \cite{Ormeno2002}, 
 $\lambda_a(0)\approx 260$~nm \cite{Chia2003} and $\lambda_a(0)=281(14)$~nm \cite{Ozcan2003}. These experiments were performed in an extremely low magnetic field ($< 10 \mu$T) in the Meissner state. 
 %were the two SC gaps of \CeCoIn{} are open. In contrary, in the present experiment, the applied field is larger than the field required to close the smallest gap $\mu_0H_{c2\, S}<10$ mT \cite{Seyfarth2008}.
%With two open gaps $n_S$ is larger in the surface impedance measurements which explain the observation of a smaller $\lambda_a(0)$.
%These differences 
%which can explain the observation of a smaller $\lambda_a(0)$.

\begin{figure}[htb]
\includegraphics[width=0.5\textwidth]{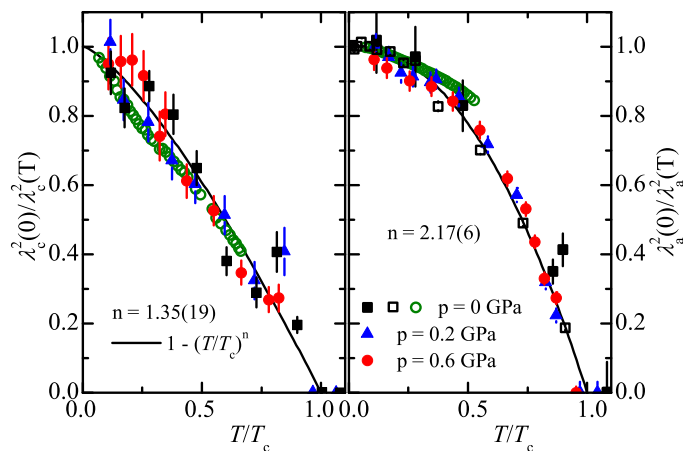}%
\caption{\label{Tdep}(Color online) $\lambda_i^2(0)/\lambda_i^2(T)$ for the two main orientations $i=a,c$ for \CeCoIn . Interestingly, $\lambda_i^2(0)/\lambda_i^2(T)$ is pressure independent in the range investigated. Full symbol were taken with the GPD spectrometer, empty squares with the LTF spectrometer. The empty green circles are 
%adapted from Ref. \cite{Chia2003}. 
obtained from tunnel-diode oscillator experiments adapted from Ref.~\cite{Chia2003}.
Black lines are power law fits to the data as described in the text.
} 
\end{figure}

\begin{figure}
\includegraphics[width=0.48\textwidth]{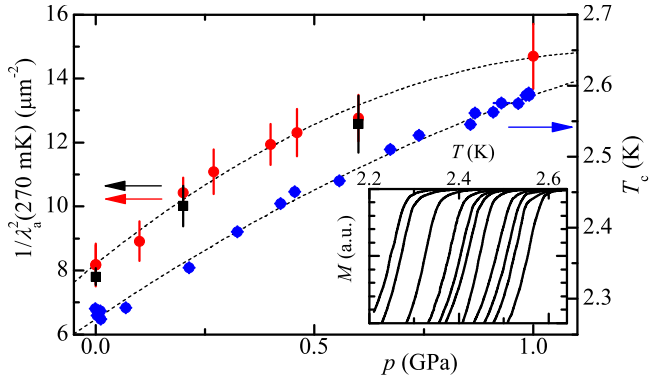}%
\caption{\label{p_dep}(Color online) Pressure dependence of $1/\lambda_a^2(270$~mK) (left scale, same symbol convention as in Fig.~\ref{Lambda_a}) measured by \muSR{} and \Tc{} (right scale) determined by SQUID magnetometry of \CeCoIn . Dashed lines are parabolic pressure dependences. Inset: magnetization curves recorded at selected pressures and used to determine \Tc{} (\Tc{} criterion is the onset of the transition).} 
\end{figure}

The temperature dependence of $\lambda_i^2(0)/\lambda_i^2(T)$ ($i=a,c$) is displayed in Fig.~\ref{Tdep}, together with a fit of the form $1-(T/T_c)^n$. The pressure evolution of \Tc{} was determined independently by SQUID magnetometry (Fig.~\ref{p_dep}). 
%The values obtained for $n$ are similar to the results obtained by other techniques, which  were interpreted either as an indication of nonlocal electrodynamics \cite{Chia2003}, gapless superconductivity \cite{Kogan2009} or an effect of the proximity to a quantum critical point \cite{Ozcan2003}.
The exponent $n$ was found to be $n=2.17(6)$ for $i=a$ in agreement with Ref.~\cite{Kogan2009}, while $n=1.35(19)$ for $i=c$. Similar temperature dependences can be obtained using $\Delta\lambda_i=\lambda_i(T)-\lambda_i(0)$ measured by tunnel-diode oscillator experiments \cite{Chia2003} taking $\lambda_a(0) \simeq 336$~nm and $\lambda_c(0) \simeq 421$~nm from this work (green empty circles in Fig.~\ref{Tdep}).
Within precision both $\lambda_i^2(0)/\lambda_i^2(T)$ ($i=a,c$) are pressure independent, suggesting that the gap symmetry is unchanged, and the value of the gap to \Tc{} ratio is constant, in agreement with the variation of less than 10\% obtained by NQR \cite{Yashima2004} in this pressure range.

The pressure independent value of $\gamma=\sqrt{m^\star_c/m^\star_a}\simeq 1.3$ (Fig.~\ref{Lambda_a}) is in good agreement with the constant value deduced from the initial slope of the upper critical field under pressure $H_{c2}'(H\perp c,p)/H_{c2}'(H\parallel c,p)\propto m^\star_c/m^\star_a\approx 2$ \cite{Knebel2010,Howald2011b}.
The same experiments indicate that the variation of $m^\star$ with pressure is less than 10\% between 0-1~GPa with a maximum around 0.5~GPa. Such a small variation cannot explain the pressure dependence of $1/\lambda_a^2$ plotted in Fig.~\ref{p_dep}. Therefore, we conclude that, within the London model [Eq. (\ref{London})],  $n_S$ increases with pressure. %The report that hole doping (Cd substitution at the In site), likely to reduce $n_S$, acts ``like'' negative pressure \cite{Pham2006} corroborate this idea.
Using an average value $m^\star_a \approx 50 m_0$ \cite{Settai2001} one obtains from Eq. (\ref{London}) a change of $n_S$ from $n_S \approx 1.8$ to $3.4$ carriers per unit cell between 0 and 1 GPa.

\begin{figure}
\includegraphics[width=0.48\textwidth]{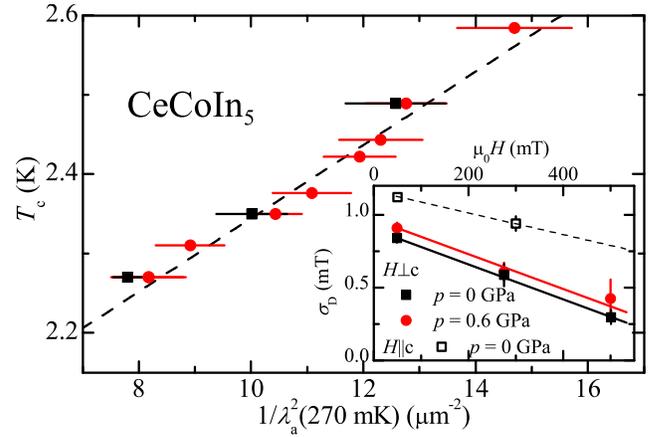}%
\caption{\label{Uemura}(Color online) Plot of $T_c$ versus $1/\lambda_a^2(270$~mK) for \CeCoIn{} under pressure (same symbol convention as in Fig.~\ref{Lambda_a}). A linear relation (dashed line $T_c(\lambda_a)=T_{c0}+A/\lambda_a^2$) with $T_{c0}=1.88(3)$~K and $A=4.6(3)\cdot 10^{-2}$ K$\mu$m$^2$  
is found. 
Inset: field dependence of the magnetic field standard deviation $\sigma_D$ (see text for details).} 
\end{figure}

%Such a strong increase of $n_S$ is certainly linked to the unusual pressure dependence of the SC state in \CeCoIn{} \cite{Howald2011b}. 
In the following we discuss two possible scenarios for this strong increase of $n_S$. The first one relies on the proposed multigap SC state of \CeCoIn{} \cite{Seyfarth2008}. The observed increase of $n_S$ with pressure would result from an increase of the small gap at 50~mT. 
Indeed, at ambient pressure for $\mu_0H \approx 50$~mT the smaller gap is already closed \cite{Seyfarth2008}. To check whether the small gap could open under pressure, we probed the field dependence of the total magnetic field standard deviation in the sample $\sigma_D$. Here $\sigma_D^2=\sigma_S^2+(0.06092\Phi_0/\lambda^2)^2$ \cite{Yaouanc2011} is the quadratic sum of $\sigma_S$ previously defined and the magnetic field standard deviation generated by the VL ($\Phi_0$ is the flux quantum). 
For comparison, for the two-gap superconductor PrOs$_4$Sb$_{12}$ \cite{Seyfarth2005} different slopes $d\sigma_D/dH$ are observed between the low magnetic field regime with two opened SC gaps and the high magnetic field regime where only one SC gap is present \cite{MacLaughlin2002}. 
In \CeCoIn , the fact that the magnetic field dependence of $\sigma_D$ is the same at 0~GPa and 0.6~GPa (inset Fig.~\ref{Uemura}), strongly suggests that a single gap is probed in the full pressure range.

Another scenario is based on an increase of the Ce valence (orbital occupancy $n_V \approx 0.9$ \cite{Booth2011} at ambient pressure). Such a scenario was proposed for the parent compound CeRhIn$_5$ where a similar decrease of $\lambda_a$ is observed between $p= 2.07$~GPa and $p= 2.26$~GPa \cite{Heffner2010}. Note that valence fluctuations are often associated with SC in Ce based heavy fermions \cite{Watanabe2011}. %The decrease of the Sommerfeld coefficient ($C/T$) of a factor 3/2 in this pressure range \cite{Sparn2002} while the Fermi velocity ($v_F$) stay constant \cite{Howald2011b} suggest indeed a decrease of the number of heavy quasiparticles ($n_q$) of a factor 1.8 ($C/T\propto n_q^{2/3}/v_F$ in a free electron gas model).

An interesting observation is the linear relation $T_c(\lambda_a)=T_{c0}+A/\lambda_a^2(270\text{~mK})$ shown in Fig.~\ref{Uemura}. This relation has some analogy with the Uemura plot [$T_c(\lambda_a)=A_{\rm U}/\lambda_a^2(0)$] \cite{Uemura1991} found for underdopped cuprate superconductors and other electronically doped unconventional superconductors. 
However, substantial differences exist: (i) no proportionality ($T_{c0}\neq 0$) and (ii) $A$ is about 90 times smaller than $A_{\rm U}$.   
In addition, pressure affects $1/\lambda_a^2$ much more in \CeCoIn{} than in cuprates \cite{Maisuradze2011}.

%Understanding an increase of $n_S$ with pressure in a system believed to have fully delocalized 4$f$-electrons \cite{Knebel2011} is difficult.
%A possible scenario is an increase of the Ce valence (orbital occupancy $n_V \approx 0.9$  \cite{Booth2011} at ambient pressure). As the Ce 4$f$-electrons are either part of the conduction band or form a Kondo singlet, this would not modify the Fermi volume \cite{Coleman2007}. However, the number of ``free'' charge carriers in the conduction band, possibly probed by $n_S$, would strongly increase. Indeed, for each electron going from an orbital state to the conduction band, an electron is also released from Kondo screening \cite{Coleman2007}.

In conclusion, we show by TF-\muSR{} that in \CeCoIn{} the magnetic penetration depth ($\lambda_a$) decreases under pressure, while the anisotropy ($\gamma=\lambda_c/\lambda_a$) and the temperature dependence of the penetration depth ratios $\lambda_a^2(0)/\lambda_a^2(T)$ and $\lambda_c^2(0)/\lambda_c^2(T)$ are almost unaffected. 
In the range of pressure investigated, a linear dependence between \Tc{} and $1/\lambda_a^2$ was found.
Within the London model, the decrease of $\lambda_a(270\text{~mK})$ under pressure corresponds to a doubling of the number density ($n_S$) between 0 and 1 GPa, possibly related with the presence of a quantum critical point. %This may indicate a valence transition of the Ce electrons. 

\begin{acknowledgments}
This work was performed at the Swiss Muon Source (S$\mu$S), Paul Scherrer Institut (PSI), Switzerland.
We acknowledge support by the Swiss National Science Foundation and the NCCR Program MaNEP.
\end{acknowledgments}

%\bibliography{C:/biblio}


\end{document}